\documentclass[onecolumn,showpacs,preprintnumbers,amsmath,amssymb,preprint]{revtex4}
\usepackage{graphicx}
\usepackage{dcolumn}
\usepackage{bm}
\begin{document}


\title{Modification of the Radiation of a  Luminescent dye embedded in a 
finite one-dimensional Photonic Crystal}

\author{J. Manzanares-Martinez}
\author{P. Castro-Garay}
\affiliation{Departamento de Investigaci\'on en F\'isica de la Universidad
de Sonora, Apartado Postal 5-088, Hermosillo, Sonora 83190,
M\'exico}

\date{\today}
\begin{abstract}
A numerical modeling of the radiation  emitted from a Luminescent dye
 embedded in a finite one-dimensional photonic crystal is presented.
The Photonic Band Structure and the Photonic Density of States are 
derived using classical electromagnetic approach, and the 
Finite Difference Time-Domain formalism 
is used to calculate the electromagnetic field distribution.
It is found that the periodic modulation provides an effective way to control the 
Spontaneous Emission under  certain circumstances. 
We find the conditions where a large amount of  light can be enhanced
on the vicinity of a photonic band edge due to the presence of  a high density of states. 
 This phenomena opens the possibility to design
new Lasers sources.
\end{abstract}
\pacs{Valid PACS appear here}

\maketitle

It is a long known fact that the Spontaneous Emission (SE) 
is not an immutable property of the source but can be altered by 
 the environment in which the atom is situated \cite{PhysRev.69.37}. 
The SE  in  a periodic medium such as a Photonic Crystal (PC)
 leads to many effects not present on
unbounded media \cite{PhysRevA.53.2799}. 
In particular,  the modification of the radiation
of a Luminescent dye embedded in a PC has been
verified in many experiments.\cite{polman:1,802983,802996} 
This phenomenon has considerable consequences
in both,  Science and Technology \cite{ISI:000279963000001}. In Science, because
SE is a Quantum phenomenon effect that
can be now study not only in relation to the existence of
Photonic Band Gaps (PBG) -frequency ranges in which no
propagating modes exist-  but also within the interaction
of more sophisticated phenomenas such as
nonlinearities \cite{PhysRevLett.58.2059,kita:161108}, 
cavity interactions \cite{1367-2630-12-5-053005}, etc.
For Technology, because a monochromatic  enhancement of the emitted light 
opens the possibility to design new Laser sources \cite{zhao:063103}.

The SE have been explained by Classical and Quantum Electrodynamic approaches. 
\cite{PhysRevA.53.2799}. 
The Classical Electrodynamics approach study the SE in terms of the deformation of the radiation 
field of a dipole due to the multiple interferences of a periodic infinite media.
\cite{PhysRevE.72.056609}.
The Quantum Electrodynamics describes the SE  as 
a formalism based on the restriction of discrete electromagnetic field values when 
modeling it, and at first-order perturbation theory the emission rate can be calculated using 
Fermi's golden rule
\cite{PhysRevA.41.2668}. In both approximations
the formulation is based on the unrealistic  consideration of an infinite crystal.
A practical realization of SE on PC is inevitably of finite size,
however. On the other hand, the consideration of an dipole as the source of radiation is also unrealistic.
Dipolar radiation produces radiation with an intrinsic 
spatial distribution that does not occur on Luminescence experiments.\cite{PhysRevE.68.036608}
The luminescent dye as a source of electromagnetic field is more likely to a 
Gaussian source with a central field frequency than the radiation field of a dipole.\cite{PhysRevE.72.017601}
 One has to note that that all kind of approximations, Classical
or from Quantum theory have to establish its connection with the experimental conditions. 
In this case, the more suitable experiment to understand the Luminescence in
presence of PBG materials
is to study  the SE modification of a Gaussian source   immersed in a finite PC.

The PBG materials are  periodic structures which do not 
allow propagation of photons over a finite band of frequencies \cite{PhysRevLett.58.2059}.
 A one-dimensional (1D)  PBG structure is the simplest PC  and usually is called 
Bragg Reflector when restricted to a quarter-wave stack.\cite{Garmire:03}
1D-PBG structures
have  interesting effects as anomalous group velocity \cite{SV_21_18}
and dispersion \cite{RMF_54_95},  existence of giant PBG 
\cite{JEMWA_24_351,PIER_111_105}
and the tunable control of the band gap edges \cite{PhysRevB.72.035336,manzanares-martinez:101110}.

The control  of fluorescence have been predicted on several PBG \cite{PhysRevA.65.043808}.
In the present study, we investigate the change on the emitted power spectra  
due to the placement of a Luminescent dye within a finite 1D-PC.
We have recently considered this problem theoretically
 to determine the degree of tuning of lasing due to 
the variation of the ambient condition in a liquid crystal based PC \cite{JEMWA_24_1867}.
The enhancement and suppression of the spontaneous emission have also been predicted 
and experimentally verified in higher dimensional PBG structures \cite{ursaki:1001,gaponik:1029}.

 In this work, we perform numerical modeling by using the
Finite Difference Time-Domain method (FDTD) which is a
methodology  closer to the experimental conditions than
the analytical formulations \cite{PhysRevA.41.2668,PhysRevE.72.056609}.
We solve numerically the Maxwell equations using FDTD simulations performed by Meep, 
a freely available software package.\cite{meep}
 This approach is flexible and also opens the possibility  to account of defects  or surface
states \cite{PhysRevB.66.113101}.

\begin{figure}
\includegraphics[scale=0.8,clip]{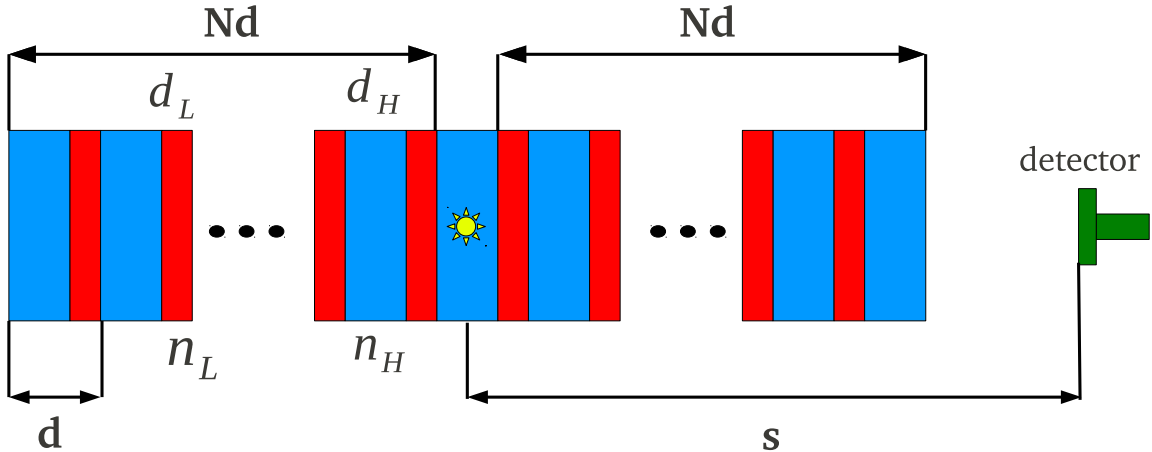}
\caption{\label{Fig_1} Geometry of the problem. A Gaussian source
is embedded in the center of an  slab of high refractive index. The
slab is bound with an arbitrary  number of N periods at both sides.}
\end{figure}

The geometry of the problem is shown in Fig. 1. A Gaussian source 
is embedded in a finite multilayer composed by alternative layers
of high ($n_H$) and low ($n_L$) refractive indices illustrated in blue
and red, respectively. The thickness of the slabs of high and low 
refractive indices are 
$d_H$ and $d_L$. The period of the structure of the unit cell is 
$d=d_H+d_L$. The layers stacks above and below the source have an identical 
  number of $N$ periods. The total thickness of the structure is
$L=2Nd+d_H$. In the direction of the periodicity 
a  detector is placed a distance $s$ from the source.

 The dispersion relation of an infinite 1D-PC is obtained using a well 
known formula \cite{Yariv:77}

\begin{equation}
 k(\omega)d = cos(k_H d_H)cos(k_L d_L)-\frac{1}{2}
\left(
\frac{k_H}{k_L}+\frac{k_L}{k_H}
\right)
sin(k_H d_H)sin(k_L d_L)
\end{equation}

where $k(\omega)$ is the Bloch wave vector, and $k_H=n_H\omega/c$ and
$k_L=n_L\omega/c$ are the wave vectors for the high and low dielectric media, respectively.
It is expected that the radiation of the Luminescent dye embedded in the
1D-PC can be related to the  Density of States (DOS) which is defined as $\rho(\omega)$. 
$\rho(\omega)$ is obtained by  differentiating the dispersion relation in the form 
\cite{PhysRevA.46.612}

\begin{equation}
 \rho(\omega) = \frac{dk(\omega)}{d\omega}
\end{equation}

The DOS is  the number of allowed states and is obtained 
taking the  relation between  the number of modes $\Delta k$ available for photons 
within the frequency $\Delta \omega$.

\begin{figure}
\includegraphics[scale=1.0,clip]{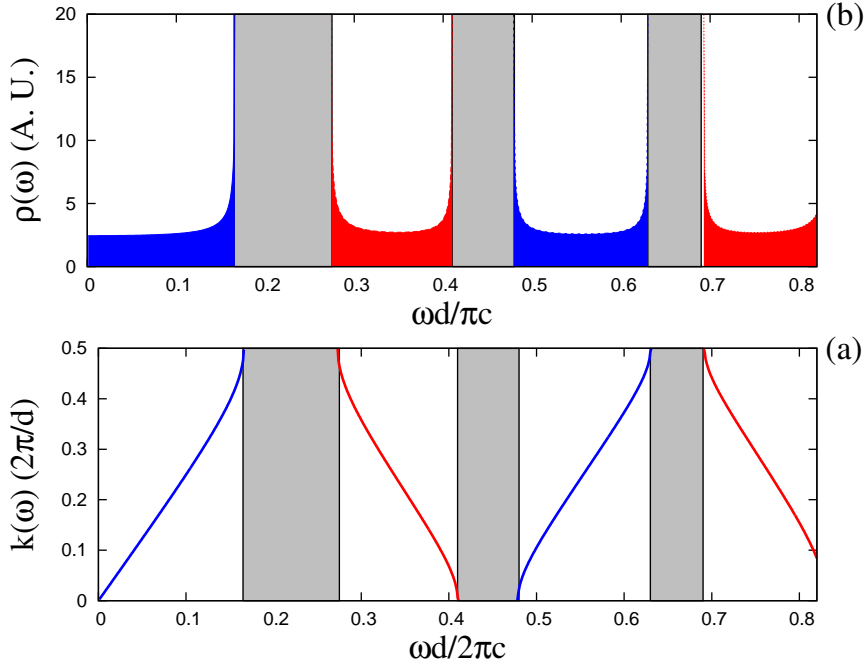}
\caption{\label{Fig_2}  Relation dispersion and DOS in panels (a) and (b) for
an infinite 1D-PC of high and low refractive indices $n_H=3.42$ and $n_L=1.49,$
respectively. The width of the layers is $d_H=0.4d$ and $d_L=0.6d$, in each case.}
\end{figure}

A particular case of the relation dispersion and DOS is shown in Fig. 2 in panels (a) and (b), 
respectively. We take the values of the indices as $n_H=3.42$ and $n_L=1.49$. The
width of the layers are $d_H=0.4d$ and $d_L=0.6d$. We observe in panel (a) the existence 
of four bands and three PBG. It has been reported by 
Joannopoulous {\it et al.} that the bands can be distinguished by where
the power of the modes lies \cite{PC_Joannopoulous}. The energy of the bands can be concentrate
in the high index media or in the low index media. For this reason
we refer to ''high dielectric band'' and ''low dielectric bands''. A similar situation
exist in electronic band structure of semiconductors where are defined the ''conduction''
 and  ''valence'' bands \cite{PhysRevLett.67.3380}.
We present in blue color the first and the third bands 
which are related to electromagnetic modes located in the the high index media. 
In the same manner, the second and the
fourth bands are related to a electromagnetic mode located in the low index media.
In panel (b) we present the DOS for each band that is obtained from the derivative of the 
relation dispersion using eq. (2). We observe an high value of the DOS at 
all the band edges. For clarity, we present in gray color all the PBG regions.

\begin{figure}
\includegraphics[scale=1.0,clip]{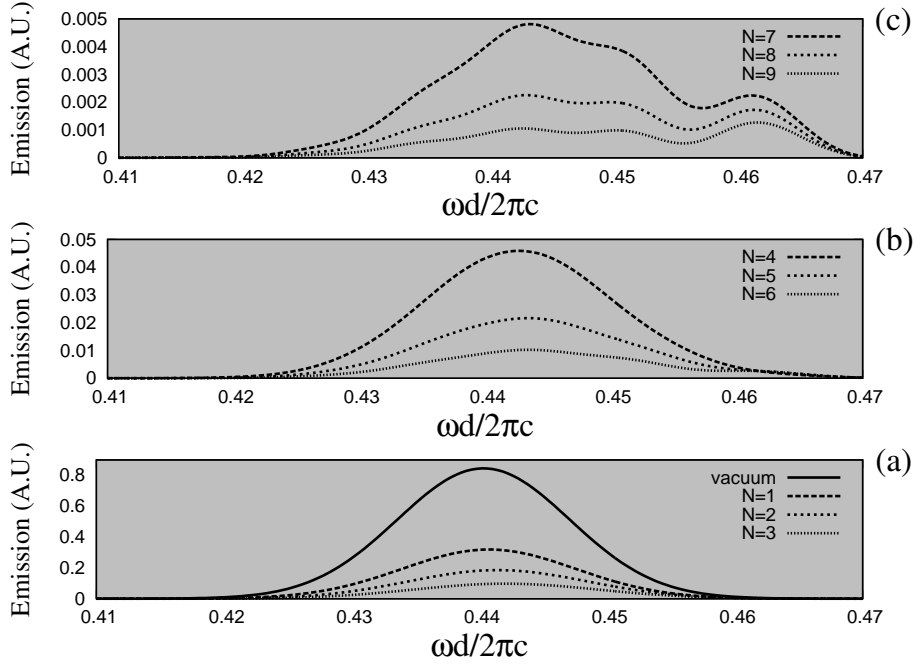}
\caption{ SE from a Gaussian source embeeded in a finite 1D-PC. The source is 
emitting at the center of the PBG with a central frequency $\omega_c=0.44$
and a width of $\sigma=0.06$. In panel (a) we present with solid line 
the emission for  the case of the vacuum and for the cases of N=1,2,3 with dashed, 
dotted and tiny dotted, respectively. In panels (b) and (c)
we present the cases of N=4,5,6 and N=7,8,9.}
\end{figure}

The SE  a Luminescent dye located in the center of the finite 1D-PC
is modeled  as a Gaussian source for the electric field ($E_s$) in 
the form \cite{PhysRevE.72.017601}

\begin{equation}
 E_s(\omega) = E_0 e^{-0.5 [{\omega-\omega_c}/{\sigma}]^2}
\end{equation}

where $\omega_c$ and $\sigma$ are the central frecuency and width of the pulse, respectively.

The SE of a Gaussian source embedded in a Finite 1D-PC are displayed in Fig. 3.
We consider that the source have a central frequency of $\omega_c=0.44$ and  $\sigma=0.06$,
respectively.
In panel (a) we present as a reference the case of the emission in the vacuum 
with a solid line. We also present the cases of $N=1,2,3$. In panels (b) and (c) 
we present the emission for $N=4,5,6$ and $N=7,8,9$. We observe that as we increases 
the number of layers, the SE decreases. This is the expected results because the
source is emitting in the PBG.

\begin{figure}
\includegraphics[scale=1.0,clip]{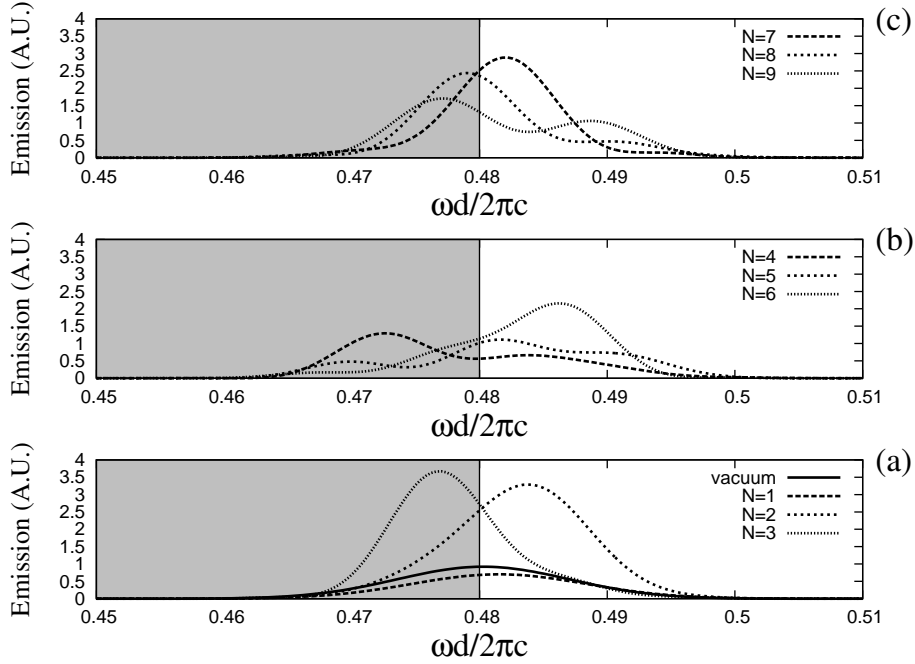}
\caption{SE from a Gaussian source embeeded in a finite 1D-PC. The source is 
emitting at the center of the PBG with a central frequency $\omega_c=0.48$
and a width of $\sigma=0.06$. In panel (a) we present with solid line 
the emission for  the case of the vacuum and for the cases of N=1,2,3 with dashed, 
dotted and tiny dotted, respectively. In panels (b) and (c)
we present the cases of N=4,5,6 and N=7,8,9.}
\end{figure}

In Fig. 4 we present the cases of the SE at the high energy limit of the band gap.
We take a central frequency and pulse width of $\omega_c=0.48$ and $\sigma=0.06$, respectively.
We observe in panel (a) that the SE for the case of the vacuum and $N=1$
 is almost similar, but it exist a dramatic enhancement of the radiation for 
$N=2,3$. This is result of a redistribution of the electromagnetic radiation
induced by the existence of allowed Bloch modes. This preferential distribution of the
electromagnetic field in the periodicity direction is similar in panels (b)and (c).

\begin{figure}
\includegraphics[scale=1.0,clip]{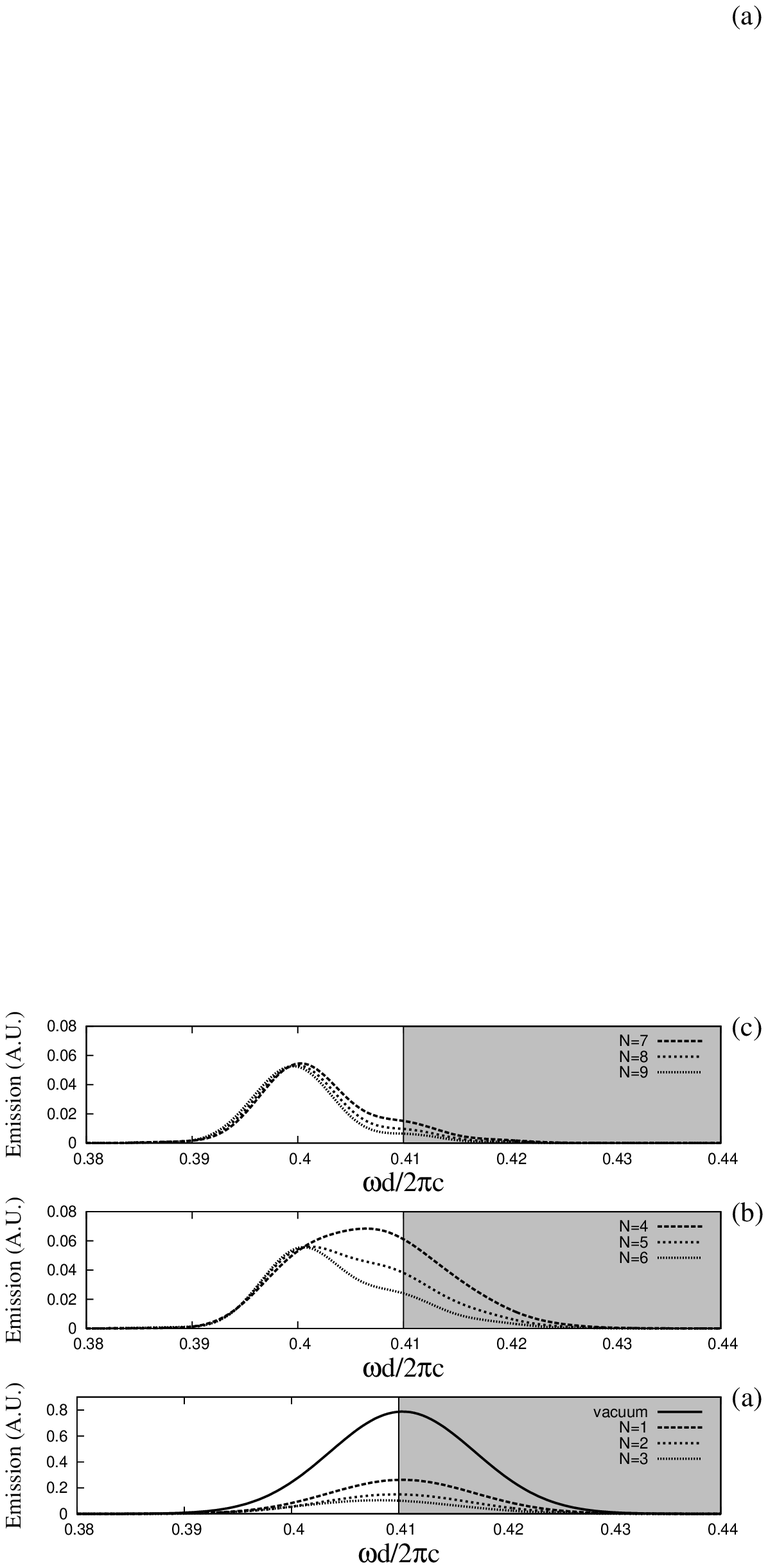}
\caption{SE from a Gaussian source embeeded in a finite 1D-PC. The source is 
emitting at the center of the PBG with a central frequency $\omega_c=0.41$
and a width of $\sigma=0.06$. In panel (a) we present with solid line 
the emission for  the case of the vacuum and for the cases of N=1,2,3 with dashed, 
dotted and tiny dotted, respectively. In panels (b) and (c)
we present the cases of N=4,5,6 and N=7,8,9.}
\end{figure}

The SE emission at the low energy limit of the band gap
is presented in Fig. 5. Here we consider the values
 $\omega_c=0.41$ and $\sigma=0.06$,respectively.
In difference to the case of the high energy limit,
here we do not observe an enhancement of the radiation measured by the detector,
as it could be expected by the presence of the band gap edge. We observe in panel (a)
a decreasing in the radiation. In panels (b) and (c) we observe a kind of stabilization of the 
radiated power.

\begin{figure}
\includegraphics[scale=1.0,clip]{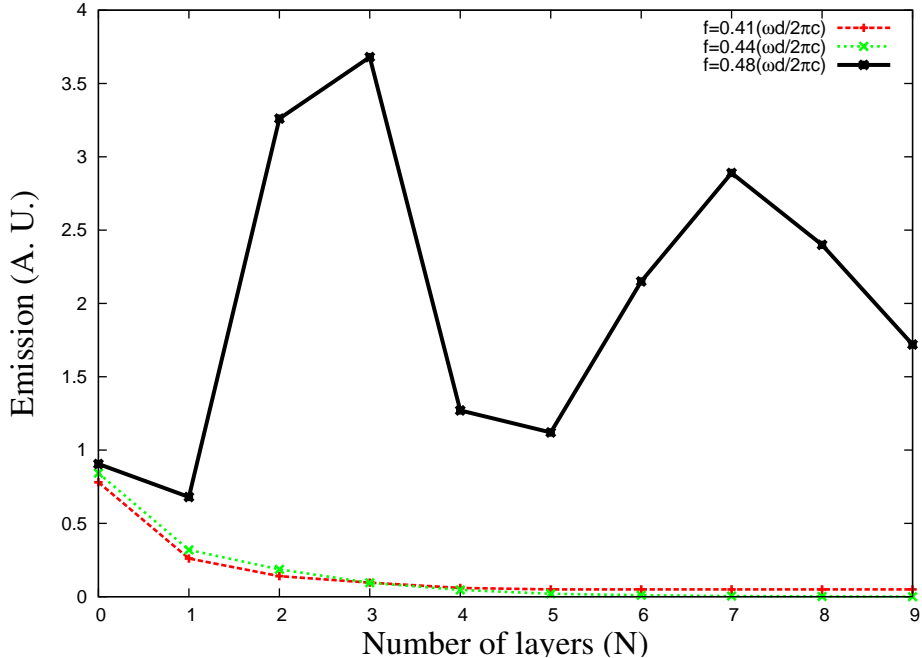}
\caption{Comparison of the Emission as a function of the number of layers N for the
cases $\omega_c=0.41$, $\omega_c=0.44$ and $\omega_c=0.48$ displayed with red dotted,
green tiny dotted and  black solid, respectively. }
\end{figure}

A global vision of the emitted radiation for the tree cases can be 
observed in Fig. 6. For the cases of the emission at the middle of the PBG ($\omega_c=0.44$)
 and at the lower band edge ($\omega_c=0.41$),
we observe an important decreasing of the emission.
In contrast, at the upper PBG ($\omega_c=0.48$) 
we observe a high enhancement of the radiation.
However, it is strange the low coupling of the electromagnetic field 
at the low energy limit of the band gap. 
Why the electromagnetic field is not coupled with this band?

\begin{figure}
\includegraphics[scale=0.7,clip]{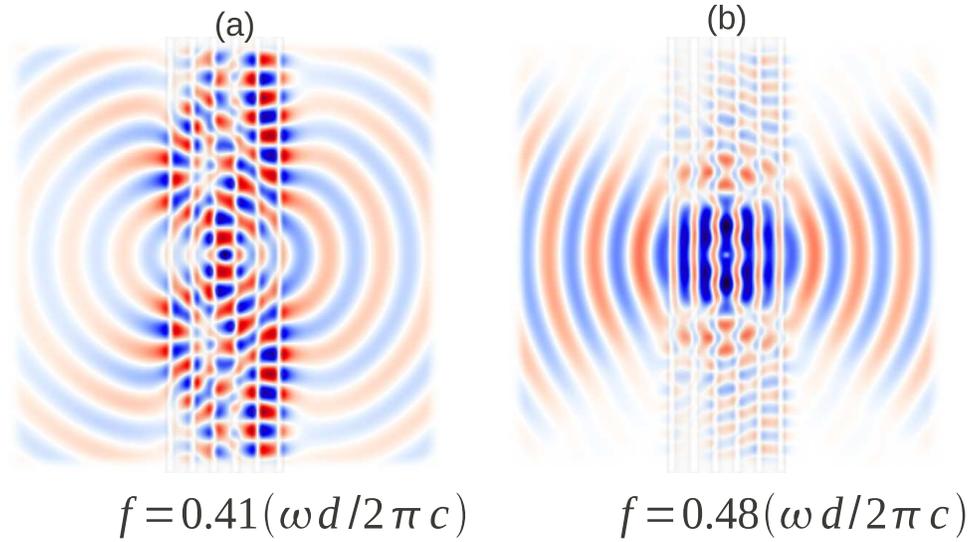}
\caption{Electromagnetic field at the band gap edges. In panel (a) and (b) 
we have  the field distribution for the low  and high limits of the band gaps, respectively.}
\end{figure}

In order to explain this difference, we show in Fig. 7 the electromagnetic field
produced at  the low ($\omega_c=0.41$) and high ($\omega_c=0.48$) 
band edges  in panels (a) and (b), respectively.
We observe that in panel (a) exist a complicate mode, where a 
destructive interference exist that does not allow the 
excitation of the low index mode. 
In contrast, in panel (b) we observe that the 
high index mode is excited. The reason of this difference is that the
source of electromagnetic field is placed at the high index material.

In conclusion, we have presented a numerical modeling of the emission of
a Luminescent dye embedded in a finite 1D-PC. The treatment allows us to understand
 that is possible
the enhancement of the radiation at the high energy limit of the band gap. 
We have also demonstrated that the DOS can not be directly related to an
enhancement of the emitted power
because it is necessary to take into account the conditions of coupling between
 the radiation of the 
source to the allowed Bloch modes at the band edge.

\section{Acknowledges}

PCG thanks the Optical Academy of the Department for Physics Research of the 
Sonora University for a temporary contract as Research Associate.

\bibliography{test2}

\end{document}